\documentclass[a4]{jpsj-suppl}
\usepackage{times}
\usepackage{color}
\usepackage{bm}

\title{
Magnetic and Dielectric Properties in Multiferroic 
Cu$_{3}$Mo$_{2}$O$_{9}$ under High Magnetic Fields
}

\author{
Haruhiko {\sc Kuroe}$^{1}$,
Kento {\sc Aoki}$^{1}$, 
Ryo {\sc Kino}$^{1}$,
Tasuku {\sc Sato}$^{1}$, 
Hideki {\sc Kuwahara}$^{1}$, 
Tomoyuki {\sc Sekine}$^{1}$, 
Takumi {\sc Kihara}$^{2}$, 
Mitsuru {\sc Akaki}$^{2}$, 
Yoshimitsu {\sc Kohama}$^{2}$, 
Masashi {\sc Tokunaga}$^{2}$, 
Akira {\sc Matsuo}$^{2}$, 
Koichi {\sc Kindo}$^{2}$, 
Masashi {\sc Hase}$^{3}$, 
Kanji {\sc Takehana}$^{3}$, 
Hideaki {\sc Kitazawa}$^{3}$, 
Kunihiko {\sc Oka}$^{4}$, 
Toshimitsu {\sc Ito}$^{4}$, and 
Hiroshi {\sc Eisaki}$^{4}$ 
}
\inst{
$^{1}$ Physics Division, Sophia University, %
7-1 Kioi-cho, Chiyoda-ku, Tokyo 102-8554, Japan\\
$^{2}$ The Institute for Solid-State Physics (ISSP), the University of Tokyo, 
Kashiwa, Chiba 277-8581 Japan\\
$^{3}$ National Institute for Materials Science (NIMS),%
Tsukuba, Ibaraki 305-0047, Japan\\
$^{4}$
National Institute of Advanced Industrial Science and Technology (AIST), %
Tsukuba, Ibaraki 305-8568, Japan
}

\email{kuroe@sophia.ac.jp}

\recdate{September 30, 2013}

\abst{
 The magnetic and dielectric properties 
under high magnetic fields 
are studied in the single crystal of Cu$_{3}$Mo$_{2}$O$_{9}$.
This multiferroic compound 
has distorted tetrahedral spin chains.
The effects of the quasi-one dimensionality and 
the geometrical spin frustration 
are expected to appear simultaneously.
We measure the magnetoelectric current and 
the differential magnetization 
under the pulsed magnetic field up to 74 T.
We also measure the electric polarization 
versus the electric field curve/loop 
under the static field up to 23 T.
Dielectric properties change at the magnetic fields 
where the magnetization jumps are observed in the magnetization curve.
Moreover, the magnetization plateaus are found at high magnetic fields.
}

\kword{magnetization plateau, multiferroic material, geometrical spin frustration}

\begin{document}
\maketitle

\section{Introduction}

The multiferroic behavior, 
especially the spin-driven ferroelectricity, 
is one of the most recent topics 
in solid-state physics.
In a multiferroic phase, 
where a long-range magnetic order 
coexists with a ferroelectricity, 
both of the time reversal and the spatial inversion 
symmetries are broken.
In many spin-driven multiferroic materials, 
including the first reported material 
of this class TbMnO$_{3}$ \cite{KimuraNature2003}, 
the origin of the electric polarization is 
explained by the spin-current model \cite{Katsuura2005,Kenzelmann2005,Mostovoy2006}.
In this model, 
a magnetic superlattice formation plays an essential role 
in breaking these symmetries.
Also, the geometrical spin frustration effect, itself, 
gives a potential to cause a multiferroic behaviour.
The ground state of a perfectly frustrated system 
is characterized as a macroscopically degenerated one. 
When this degeneracy is lifted, 
the low-energy excited states appear.
These low-energy states lead 
to field-induced successive phase transitions.
The simplest example of this class is 
an antiferromagnetic (AFM) spin triangle, 
where a electric polarization 
is predicted to appear \cite{Bulaevskii2008,Khomskii2010}.

We have studied the multiferroic properties 
in Cu$_{3}$Mo$_{2}$O$_{9}$ \cite{Kuroe2011}.
This compound has the spin frustrating 
quasi-one dimensional distorted tetrahedral quantum spin system 
made from $S$ = 1/2 Cu$^{2+}$ spins 
at three different Cu sites, namely Cu1-Cu3.
The successive studies of this system were proceeded 
under the recognition that the canted AFM phase is formed 
below $T_{\rm N}$ = 8.0 K \cite{Hamasaki2008,Hamasaki2010}.
The high-quality large single crystals 
prepared with the continuous solid-state 
crystallization method enabled us 
to measure various physical quantities 
with variety of measuring techniques \cite{Oka2011}.
The results of the neutron scattering 
under zero magnetic field 
show that there is no sign of 
the magnetic superlattice formation below $T_{\rm N}$ and 
the spin excitations have two interacting branches 
of the quasi-one-dimensional spinon excitations 
mainly from the spins at the Cu1 sites 
with the dominant excitation energy of 4.0 meV 
in the chain direction and 
of the nondispersive excitations 
of the Cu2-Cu3 spin (nearly singlet) dimers 
at 5.8 meV \cite{KuroeJPCS,Kuroe2011PRB}.

The whole dispersion curves of the magnetic excitations 
at 0 T were reproduced by the bond-operator theory 
based on the picture of 
the classical spin wave of the spins at the Cu1 sites 
interacting with the Cu2-Cu3 quantum spin dimers \cite{Matsumoto2012}.
This theory predicted 
a jump of magnetization ${\bm M}$ 
accompanied by a drastic change of 
the ferroelectric polarization ${\bm P}$
in the narrow region of the magnetic field ${\bm H}$ 
($|{\bm H}| \sim$ 50 T) 
where the quantum spin dimers are magnetized.
Motivated by this prediction, 
we measured ${\bm M}$ and ${\bm P}$ 
of the single crystals of Cu$_{3}$Mo$_{2}$O$_{9}$ 
under high magnetic fields. 

\section{Experiments and Results}
\subsection{Magnetization and electric polarization under pulsed magnetic fields}
\begin{figure}[h]
\includegraphics[width=\textwidth]{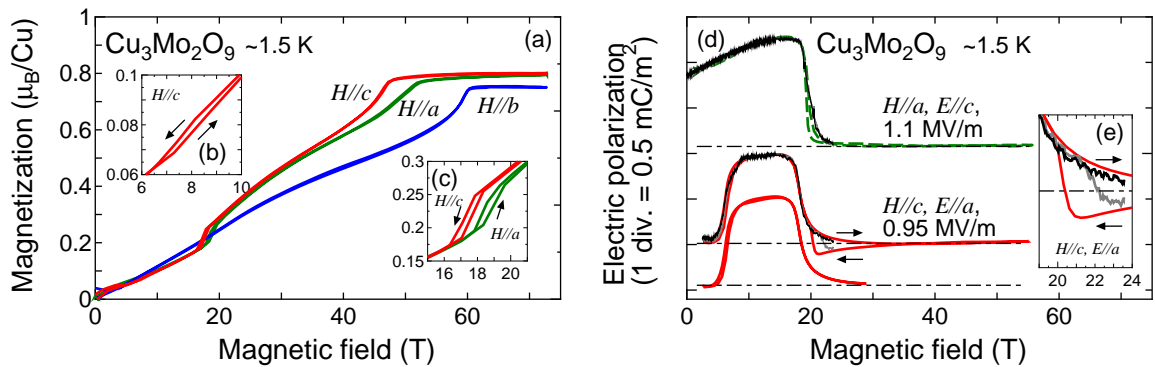}
\caption{(color online) 
The magnetization curves up to 74 T (a) and their expansions (b) and (c).
The colors distinguish the directions of magnetizations.
The electric polarizations along the $a$ ($c$) axis 
as functions of the magnetic field along the $c$ ($a$) axis 
are shown by red solid (green dashed) curves in (d) and (e).
The dot-dashed lines in (d) and (e) denote the zero electric polarization.
The black and gray curves in (d) and (e) show the 
magnetic field dependence of the electric polarization 
at a given electric field 
obtained from the electric polarization-electric field loop.
The arrows distinguish the field-increasing 
and the field-decreasing processes.
}
\label{Magnetization-Polarization}
\end{figure}
The magnetization curves ${\bm M}({\bm H})$ at 1.5 K 
in Fig. \ref{Magnetization-Polarization}(a)
were obtained from their derivatives 
by using the nondestructive pulsed magnets up to 74 T 
in the International MegaGauss
Science Laboratory, Institute for Solid State Physics (ISSP), 
the University of Tokyo.
The absolute values 
are obtained by comparison of the data with the ones 
taken under static magnetic fields \cite{Kuroe2011,Hamasaki2009}.
Hereafter, we introduce the subscript 
$x$ ($x$ = $a$, $b$, and $c$)
for the $x$-component of a vector quantity.
As shown in Figs. \ref{Magnetization-Polarization}(b) 
and \ref{Magnetization-Polarization}(c),
the phase boundaries together with the jumps of $M_{c}$ ($M_{a}$) 
at 8 and 17 T (19 T) \cite{Kuroe2011,Hamasaki2009,Okubo2010}
were reproduced well.

Unexpectedly, we observed clear magnetization plateaus 
at high magnetic fields.
The lower boundary of 
the magnetization plateaus 
in $M_{a}$, $M_{b}$, and $M_{c}$ are
52.3, 60.3, and 47.5 T, respectively.
And the values of $M_{a}$, $M_{b}$, and $M_{c}$ 
at these magnetic fields 
are 0.775, 0.757, and 0.786 in the unit of Bohr magneton $\mu_{\rm B}$, 
respectively.
These plateaus correspond to the 2/3 one, 
when we suppose the anisotropic $g$ values $(g_{a}, g_{b}, g_{c})$ 
as (2.33, 2.27, 2.36).
We consider that these $g$ values are reasonable 
because the $g$ value of a polycrystalline sample at low temperature 
($g$ = 2.52 below 8 T and 2.41 above 8 T, ref. \cite{Okubo2010})
is larger than the one at room temperature 
[$(g_{a}, g_{b}, g_{c})$ = (2.090, 2.193, 2.180) 
for single crystal \cite{Hamasaki2008} and 
$g$ = 2.11 for polycrystalline sample \cite{Okubo2010}].
The deviation of the absolute values 
is probably explained 
by the drastic changes of the low-temperature ESR spectrum; 
especially, the reduction of $g$ value above 18 T is 
caused by the magnetic-field-induced phase transition.
Further study of the magnetization curves 
and ESR spectra up to the saturation magnetic fields 
are necessary to confirm 
that these magnetic plateaus correspond to the 2/3 one.

The $P_{c}$-$H_{a}$ and $P_{a}$-$H_{c}$ curves 
in Fig. \ref{Magnetization-Polarization}(d) were 
obtained by the integration of 
the magnetoelectric current 
under a static electric field 
and the pulsed magnetic field up to 55 T 
using another pulsed magnet in ISSP.
The $P_{c}$-$H_{a}$ curve was measured 
under the static electric field of 1.1 MV/m.
At 0 T, $P_{c}$ has a finite value.
This reflects the spontaneous electric polarization 
along the $c$ axis \cite{Hosaka2012}.
The $P_{c}$ increases with increasing magnetic field, 
saturates around 15 T, and shows a rapid drop around 20 T.
The magnetic-field hysteresis effects appear 
in both of $P_{c}$ and $M_{a}$, as shown in Figs. 1(c) and 1(d).
And then, we consider that 
the changes of $P_{c}$ and $M_{a}$ occur simultaneously.

The $P_{a}$-$H_{c}$ curve was measured 
under the static electric field of 0.95 MV/m, 
slightly weaker than the case of the $P_{c}$-$H_{a}$ curve.
The rapid increase of $P_{a}$ around 7 T 
has been reported before \cite{Kuroe2011}.
The $P_{a}$ shows a rapid drop at 18 T.
These changes correspond to 
the magnetic-field-induced phase transitions 
accompanied by a jump of $M_{c}$.
The $P_{a}$-$H_{c}$ curve shows 
a unique magnetic field dependence:
We observed that the sign of $P_{a}$ is flipped only 
in the field-decreasing process after 
the maximum magnetic field 
larger than the lower boundary 
of the magnetization plateau.
The sign flipping of $P_{c}$ was not observed in 
the $P_{c}$-$H_{a}$ curve.

\subsection{Electric polarization versus electric field loop under 
static magnetic fields}

To discuss the flipping of $P_{c}$, 
the detailed analysis of the ${\bm P}$-${\bm E}$ loop is necessary.
And then, we measured it under 
the static magnetic field up to 23 T 
by using the hybrid magnet 
installed at the National Institute for Material Science.
The ${\bm P}$-${\bm E}$ loops under static ${\bm H}$ 
were obtained by using the modified Sawyer-Tower circuit 
with a measurement frequency of 1 Hz.
We numerically subtracted 
the noise from the power line 
coming through the ground loop 
and the signal above $T_{\rm N}$ 
which contained the capacitance 
coming from the sample shape and the coaxial cables.

\begin{figure}[h]
\label{Polarization}
\begin{minipage}[b]{0.40\textwidth}
\caption{The electric polarization 
$P_{c}$ ($P_{a}$) along the $c$ ($a$) axis 
versus electric field loops 
in Cu$_{3}$Mo$_{2}$O$_{9}$ 
under static magnetic field $H_{a}$ ($H_{c}$) 
along the $a$ ($c$) axis.
Parentheses denote the case of Fig. 2(b).
}
\end{minipage}
\includegraphics[width=0.60\textwidth]{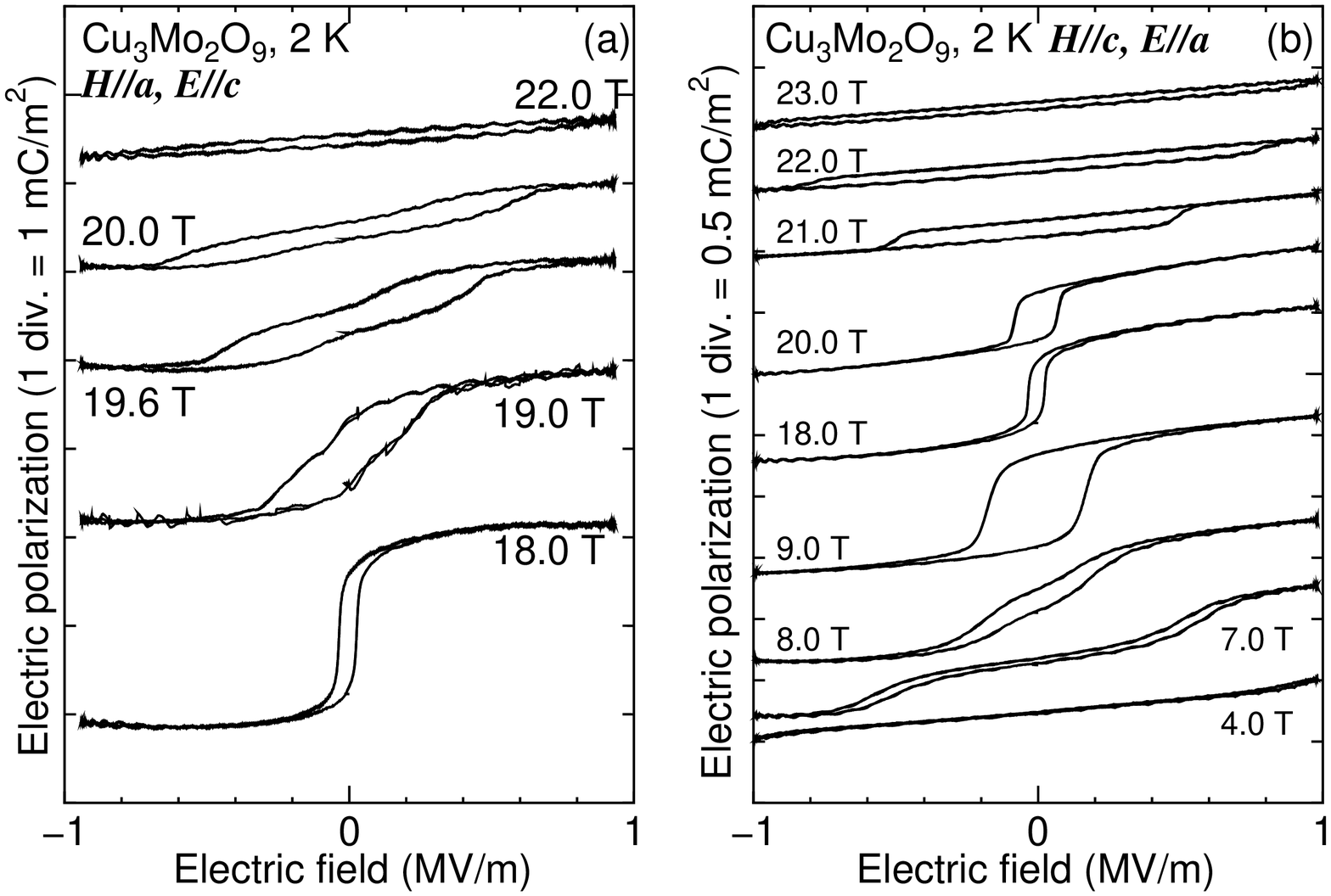}
\end{figure}
The typical data around the phase boundaries are shown 
in Figs. 2(a) and 2(b).
At $H_{a}$ $\sim$ 18 T, where the jump of $M_{a}$ is observed, 
the $P_{c}$-$E_{c}$ loop shows a drastic change:
The double-hysteresis loop is observed at 19.6 and 20.0 T.
The small loop observed at 22.0 T is 
probably a part of the double hysteresis loop.
Because the saturated electric field at 22.0 T is 
larger than the maximum value in this experiment, 
the double hysteresis loop could be observed 
as a weak and widely spread $P_{c}$-$E_{c}$ loop.

As shown in Fig. 2(b),
the $P_{a}$-$E_{a}$ loop at 4 T is (almost) closed.
The double hysteresis loop is observed at 7 T 
as a precursor of the ferroelectric transition 
around $H_{c}$ $\sim$ 8 T \cite{Kuroe2011}.
The change of the $P_{a}$-$E_{a}$ loop around 18 T 
is different from the one at 8 T; 
the coercive electric field (CEF) starts to diverge 
and the spontaneous electric polarization is suppressed.
These changes and the jump of $M_{c}$ 
characterize the magnetic-field-induced phase transition 
at 19 T.
At 23 T, the $P_{a}$-$E_{a}$ loop seems to be closed 
because the CEF is larger 
than the maximum electric field of the measurements.

To compare with the ${\bm P}$-${\bm H}$ curve 
obtained from the magnetoelectric current, 
we drew the corresponding curve 
from the $P_{c}$-$E_{c}$ ($P_{a}$-$E_{a}$) loops 
under many static $H_{a}$ ($H_{c}$) 
as the series of $P_{c}$ ($P_{a}$) 
at a given $E_{c}$ ($E_{a}$).
The typical data are shown 
by black and gray curves in 
Figs. \ref{Magnetization-Polarization}(d) 
and \ref{Magnetization-Polarization}(e).
This method has 
the advantage of detecting the metastable state.
We can distinguish 
the electric polarization 
in the electric-field increasing and decreasing processes.
As shown in Fig. \ref{Magnetization-Polarization}(d), 
the $P_{a}$-$H_{c}$ curve obtained from this procedure 
reproduces the one from the magnetoelectric current well.
However, as shown in Fig. 2(e), 
we observed a small discrepancy between them around 22 T.  
It is probably due to the scan speed of $H_{c}$ 
and slight difference of $E_{a}$.

\section{Discussion}
\subsection{Phase boundaries around 20 T}
We emphasize that 
the way how the ${\bm P}$-${\bm E}$ loop changes is important.
In the standard Landau theory of 
the temperature-induced first-order phase transition, 
as is well known, 
the free energy 
is expanded up to the sixth-order term of order parameter.
The coefficient of the squared order parameter 
plays a essential role: 
The phase transition is driven by 
a change in sign of this.
Merz introduced 
the dimensionless (electric) polarization, 
the dimensionless electric field, 
and the dimensionless temperature 
to obtain the universal curve/loop 
of the first-order ferroelectric phase transition \cite{Merz1953}.
According to this theory, 
there is a parameter region where 
the double hysteresis loop exists.

In many cases, 
the coefficient of the squared order parameter 
is treated to depend only on temperature.
However, this can depend on another physical quantity;
for example, 
this coefficient depends on pressure 
when the phase transition is driven 
by the change of lattice constant 
induced by a thermal expansion. 
In this case, the coefficient of the squared 
order parameter depends on pressure.
In some conditions the pressure-induced phase transition is possible.
In case of the magnetization of Cu$_{3}$Mo$_{2}$O$_{9}$, 
the hydrostatic pressure causes 
the change of effective temperature \cite{Hamasaki2010}.
In the present work, 
at $H_{c}$ $\sim$ 8 T and at $H_{a}$ $\sim$ 20 T, 
we observed the magnetic-field induced phase transitions 
together with the double hysteresis loop.
The origin of these phase transitions is thought to be 
a change in sign of the squared order parameter 
under the magnetic field.
One of the possible origins is 
the introduction of the term of $M_{x}^{2}$$P_{x'}^{2}$, 
which is the lowest-order cross-correlation term 
having an even parity 
for both of the spatial inversion and 
the time-reversal symmetry operations.

We did not observe 
the double hysteresis loop 
but the diverge of CEF 
through the phase transition at $H_{c}$ $\sim$ 18 T.
Moreover, two $P_{a}$-$H_{c}$ curves 
obtained from the magnetoelectric current 
and the $P_{a}$-$E_{a}$ loops 
under static magnetic fields 
show similar magnetic-field dependences to each other.
And then, 
we consider that 
the observation of 
the flipping of $P_{a}$ under $H_{c}$
is related to the observation of ${\bm P}$-${\bm E}$ loop.
As shown in Figs. 2(a) and 2(b), 
the diverge of CEF was not observed 
at other phase transitions.
Because neither the flipping of electric polarization 
nor the diverge of CEF are observed 
in conventional dielectric materials, 
we conclude that 
the phase around 20 T is not 
a conventional paraelectric one.
Hereafter, we name this phase the strong CEF phase.
The phase transition to the strong CEF phase was reported 
in the 5.0\% Zn-substituted Cu$_{3}$Mo$_{2}$O$_{9}$ \cite{Kuroe2012}.
In this case, the strong CEF phase appears at 0 T.
We consider that 
the origin of these two CEF phases is important 
to understand the multiferroic phase in Cu$_{3}$Mo$_{2}$O$_{9}$.

\subsection{Magnetization curve with magnetization plateau}
To realize the magnetization more than 1/3 of the full moment, 
the Cu2-Cu3 spin dimers should be magnetized, at least partially.
It is important to clarify 
the way how the spin dimmers are magnetized 
in order to understand the magnetization plateau.
The present results of magnetization indicate that 
the Cu2-Cu3 spin dimers start to be magnetized 
under weaker magnetic field than 
the theoretically expected value \cite{Matsumoto2012}.
We need to introduce an additional term 
into the magnetic Hamiltonian
to stabilize the magnetization plateau.
It may be explained 
by the theory in the three-dimensional 
geometrically frustrated AFM pyrochlore lattice.
The biquadratic term added into the standard 
bilinear Heisenberg AFM Hamiltonian 
induced by spin-lattice coupling
leads to the magnetization plateau \cite{Penc2004,Penc2007}.

The lower boundaries of the magnetization plateaus 
in $M_{a}$, $M_{b}$, and $M_{c}$ 
should be the phase boundaries of 
the quantum phase transition induced by the external magnetic field 
because the magnetic excitation gap should open 
in the regions of magnetization plateau.
As the examples of magnetic structures with the 2/3 magnetization, 
based on the picture of valence bond solid, 
one can imagine the following:
(i) All of the spins at the Cu1 spin-chain sites and 
a half of the Cu2-Cu3 spin dimers are fully magnetized ($S$ = 1)
and the rest of the Cu2-Cu3 spin dimers stays singlet ($S$ = 0), 
i.e., the Wigner crystal of $S$ = 1 magnetically excited states 
(the triplon state) is formed;
(ii) All of Cu2-Cu3 spin dimers are fully magnetized ($S$ = 1) 
and all of the spins at the Cu1 sites are dimerized 
to become nonmagnetic ($S$ = 0).
In the model (ii), 
the lattice distortion coming from 
the strong spin-lattice coupling is necessary.
This effect is known as the spin-Peierls transition \cite{Pytte1974}, 
which is observed in the inorganic quantum spin system \cite{Hase1993}.
The study on the detailed magnetic structures 
under the magnetization plateau region is strongly desired.

\subsection{Phase diagram}
\begin{figure}[b]
\begin{minipage}[b]{0.3\textwidth}
\caption{(color online) 
The phase diagrams of 
the magnetic field along the $a$ axis (a) 
and the $c$ axis (b).
Symbols distinguish 
the experimental methods 
to obtain the phase boundary.
The solid and the dashed curves are 
only guides for the eye.
}
\end{minipage}
\includegraphics[width=0.7\textwidth]{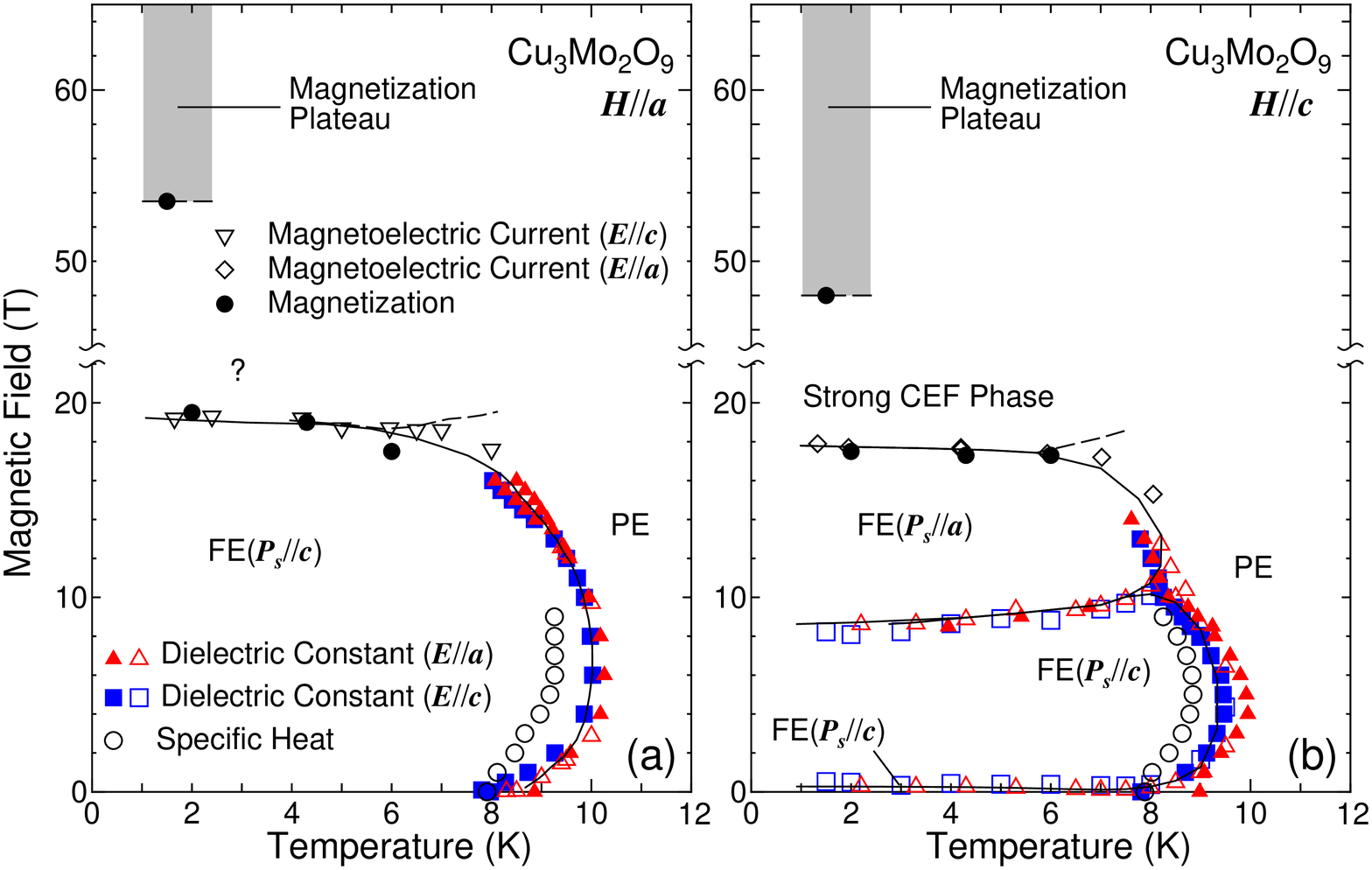}
\label{H-T_Phase_Diagram}
\end{figure}
We show the $H_{c}$-$T$ and $H_{a}$-$T$ phase diagrams 
with the phase boundaries obtained 
from the present experiments, of which detailed results are not shown here, 
and our previous ones in refs. \cite{Kuroe2011,Hosaka2012} 
in Figs. 3(a) and 3(b), respectively.
One can see from Fig. 3(a) that 
the phase boundaries just below 20 T 
obtained from the magnetic-field dependences 
of the magnetization and the electric polarization 
in this work and ref. \cite{Hamasaki2010}
connect to the ones around 8 K 
obtained from the temperature dependence of 
the specific heat \cite{Hamasaki2009} and 
the dielectric constant \cite{Kuroe2011, Hosaka2012}.
And consequently, 
we settle the regions of multiferroic phases.

The $P_{a}$-$H_{c}$ curve indicates that 
the phase above 20 T is not the paraelectric phase 
but the strong CEF phase.
The phase boundary between 
this phase and the paraelectric phase at high temperatures 
has not been observed yet.
At present, 
as well as in the $H_{c}$-$T$ phase diagram, 
%as an analogy of this phase boundary, 
we expect that 
there is a similar phase boundary in the $H_{a}$-$T$ one.
It should be confirmed experimentally.
Very recent thermodynamical measurements 
under the pulsed magnetic field clarified  
this phase boundary \cite{Kohama2013RSI}.

The present work is 
an interim report of 
the multiferroic properties of Cu$_{3}$Mo$_{2}$O$_{9}$
under high magnetic fields.
The detailed phase boundaries above 20 T, 
including the lower boundary of 
the magnetization plateau at different temperatures, 
and the magnetization curves up to 
the saturated magnetic field 
are interesting topics 
which should be clarified in future.

\section*{Acknowledgement}
We acknowledge 
Prof. S. Mitsuda in Tokyo University of Science 
for his technical supports and 
Dr. N. Terada in NIMS for helpful discussion. 
This work was supported by 
Grants-in-Aids for Scientific Research (B, No. 23340096) and (C, No. 22540350) 
of The Ministry of Education, Culture, Sports, Science, and Technology, Japan.

\end{document}